\documentclass[journal]{IEEEtran}

\usepackage{amsmath}
\usepackage{amsfonts}
\usepackage{amssymb}
\usepackage{amsthm}
\usepackage{tabularx}
\usepackage{mathrsfs} 

\usepackage{algorithm}
\usepackage{algorithmic}

\usepackage{url}

\usepackage{stfloats}
\usepackage{float}
\usepackage{graphicx}
\usepackage{xcolor}
\usepackage{subfigure}
\usepackage[colorlinks=true, allcolors=blue]{hyperref}

\makeatletter
\def\blfootnote{\xdef\@thefnmark{}\@footnotetext}
\makeatother

\begin{document}

\title{\LARGE FIRES: Fluid Integrated Reflecting and Emitting Surfaces} 
\author{Farshad~Rostami~Ghadi,~\IEEEmembership{Member},~\textit{IEEE}, 
            Kai-Kit~Wong,~\IEEEmembership{Fellow},~\textit{IEEE}, 
            Masoud~Kaveh, \IEEEmembership{Member},~\textit{IEEE}\\
            F. Javier~L\'opez-Mart\'inez,~\IEEEmembership{Senior~Member}, 
            Chan-Byoung~Chae,~\IEEEmembership{Fellow},~\textit{IEEE},\\ and
            George C. Alexandropoulos,~\IEEEmembership{Senior~Member,~IEEE}
            \vspace{-8mm}
	}
\maketitle 

\begin{abstract}
This letter introduces the concept of fluid integrated reflecting and emitting surface (FIRES), which constitutes a new paradigm seamlessly integrating the flexibility of fluid-antenna systems (FASs) with the dual functionality of simultaneous transmitting and reflecting reconfigurable intelligent surfaces (STAR-RISs). The potential of the proposed metasurface structure is studied though an FIRES-enabled multicast system based on the energy splitting protocol. In this model, the FIRES is divided into non-overlapping subareas, each functioning as a `fluid' element capable of concurrent reflection and transmission and changing its position of radiation within the subarea. In particular, we formulate an optimization problem for the design of the triple tunable features of the surface unit elements, which is solved via a tailored particle swarm optimization approach. Our results showcase that the proposed FIRES architecture significantly outperforms its conventional STAR-RIS counterpart.
\end{abstract}

%\vspace{-3.5ex}

\maketitle
%\blfootnote{\noindent Copyright (c) 2015 IEEE. Personal use of this material is permitted. However, permission to use this material for any other purposes must be obtained from the IEEE by sending a request to pubs-permissions@ieee.org.} 
\blfootnote{The work of F. Rostami Ghadi is supported by the European Union's Horizon 2022 Research and Innovation Programme under Marie Sk\l odowska-Curie Grant No. 101107993. 
%The work of M. Kaveh is supported in part by the Academy of Finland under Grants 345072 and 350464. 
The work of K. K. Wong is supported by the Engineering and Physical Sciences Research Council (EPSRC) under Grant EP/W026813/1. 
The work of F. J. L\'opez-Mart\'inez is funded by MICIU/AEI/10.13039/50110001103 and FEDER/UE through grant PID2023-149975OB-I00 (COSTUME).} 
\blfootnote{\noindent F. Rostami Ghadi and F. J. L\'opez-Mart\'inez are with the Department of Signal Theory, Networking and Communications, Research Centre for Information and Communication Technologies (CITIC-UGR), University of Granada, 18071, Granada, Spain (e-mail: $\rm {f.rostami, fjlm}@ugr.es)$.}
\blfootnote{\noindent K. K. Wong is affiliated with the Department of Electronic and Electrical Engineering, University College London, Torrington Place, WC1E 7JE, United Kingdom and he is also affiliated with Yonsei Frontier Lab, Yonsei University, Seoul, Korea (e-mail: $\rm kai\text{-}kit.wong@ucl.ac.uk$).}
\blfootnote{\noindent M. Kaveh is with the Department of Information and Communication Engineering, Aalto University, Espoo, Finland (e-mail: $\rm masoud.kaveh@aalto.fi$).}
\blfootnote{C. B. Chae is with Yonsei Frontier Lab, Yonsei University, Seoul, Korea (e-mail: $\rm cbchae@yonsei.ac.kr$).}
\blfootnote{G. C. Alexandropoulos is with the Department of Informatics and Telecommunications, National and Kapodistrian University of Athens, 16122 Athens, Greece and the Department of Electrical and Computer Engineering, University of Illinois Chicago, IL 60601, USA (e-mail: $\rm alexandg@di.uoa.gr$).}

\blfootnote{Corresponding author: Kai-Kit Wong.}
%\IEEEpeerreviewmaketitle
\vspace{0mm}

\vspace{-2mm}
\section{Introduction}\label{sec-intro}
\IEEEPARstart{R}{econfigurable} intelligent surfaces (RISs) constitute one of the key enablers of sixth-generation (6G) wireless networks owing to their ability to passively manipulate electromagnetic waves, enhancing coverage, link quality, and energy efficiency with low power consumption~\cite{wei2021channel,Basar-2024}. However, conventional RIS hardware architectures are primarily designed for unidirectional reflection-only operation~\cite{Alexandropoulos-2021}. To enable full-space coverage, simultaneously transmitting and reflecting RISs (STAR-RISs) were recently introduced \cite{mu2022sim}. Nevertheless, the dual role of transmission and reflection demands more degree of freedom (DoF) to be available to perform well in dynamic wireless environments~\cite{He-STAR-RIS-2023}.

Independently, interest is growing for using reconfigurable antenna technologies to liberate the physical layer of wireless communications. Leading this attempt is the concept of fluid antenna system (FAS) proposed by Wong {\em et al.}~in  \cite{wong2020performance,wong2021fluid} that poises to obtain more DoF for wireless systems by exploiting shape and position reconfigurability in antennas \cite{new2024tut}. Recently, research efforts have also been made to combine FAS with RIS in which FAS-equipped users were considered in an RIS-aided communication system, e.g., \cite{ghadi2024on,yao2024framework,zheng2024unlocking,ghadi2024secrecy,yao2024fas,ghadi2025phase}. Moreover, there is a new trend that introduces the FAS concept directly into the design of RIS, referred to as fluid RIS (FRIS) \cite{ye2025joint,salem2025first,xiao2025fluid}. In this FRIS model, each element on the metasurface is regarded as a `fluid' element\footnote{Like in any FAS, a `fluid' element does not necessarily involve a fluidic radiator and can be realized using electronically reconfigurable structures.} capable of not only changing the phase of its reflection coefficient, but its position, i.e., the position of reflection.

Given these prior successes, this motivates us to investigate if the FAS concept can greatly lift the performance of STAR-RISs. In particular, we introduce fluid integrated reflecting and emitting surfaces (FIRES), a novel architecture that combines the dual-mode functionality of STAR-RIS with the physical reconfiguration of fluid metasurfaces. Unlike the static STAR-RIS and the FRIS model, FIRES integrates `fluidic' elements directly into the metasurface by enabling dynamic position reconfiguration, where each sub-region of the surface consists of a reconfigurable element capable of concurrently \textit{reflecting and transmitting} signals. These elements also can reconfigure their positions of radiation within a predefined area, allowing the metasurface to adapt its spatial distribution in real time. The contributions of this letter are summarized as follows: 
\begin{itemize}
\item We first introduce the FIRES framework in the downlink and then formulate an effective rate maximization problem to jointly optimize the positions, control coefficients, and operating modes of the FIRES unit elements. 
\item To efficiently solve the resulting non-convex problem, we design a particle swarm optimization (PSO) approach. 
\item The potential of the proposed FIRES system is evaluated by means of computer simulations, with the obtained results showcasing that the FIRES consistently outperforms the conventional STAR-RIS by harnessing both electromagnetic and spatial reconfigurability, thereby enhancing connectivity, spatial reuse, and spectral efficiency in dynamic wireless environments.
\end{itemize}

\vspace{-2mm}
\section{System and Channel Models}\label{sec-sys}
\begin{figure}[!t]
\centering
\includegraphics[width=0.85\columnwidth]{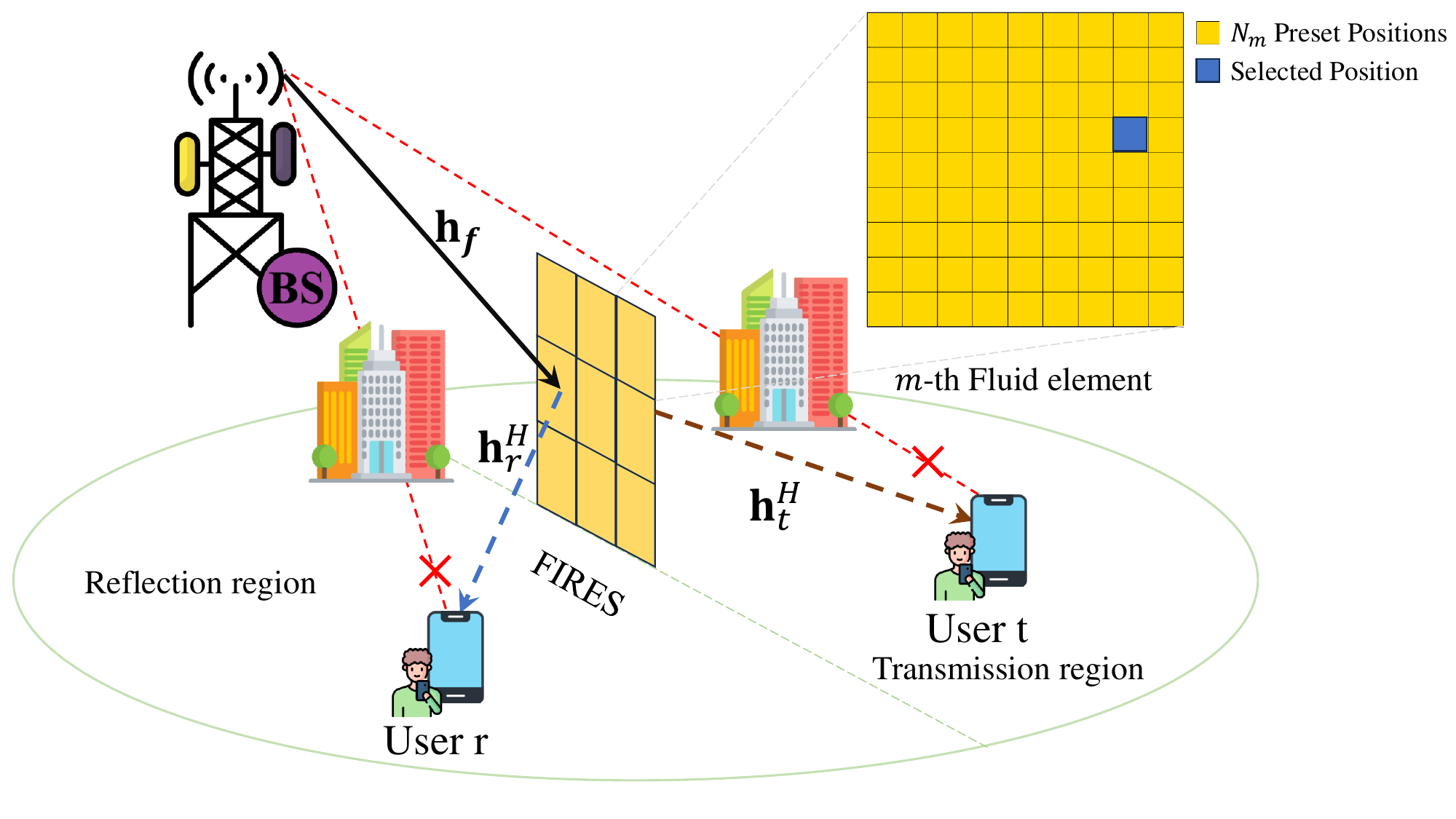}
\caption{The considered FIRES-aided communication system.}\label{fig_system}
\vspace{-3mm}
\end{figure}

We consider an FIRES-enabled communication system, as illustrated in Fig.~\ref{fig_system}, in which a fixed-position antenna (FPA) base station (BS) communicates with two FPA users $u \in \{{r,t}\}$ through an FIRES composed of fluid elements. User $r$ is located in the metasurface's reflection area, while user $t$ in the transmission one. The direct communication link between the transmitter and users is assumed blocked due to obstacles, necessitating the use of the FIRES to establish a communication path. The entire radiating surface of the FIRES is assumed to cover an area $A = A_h \times A_v\,\mathrm{m}^2$ which is divided into $M$ non-overlapping subareas. Each subarea represents a two-dimensional (2D) planar fluid element so that the FIRES consists of a total of $M$ fluid elements. 

The position of the $m$-th fluid element is characterized by the coordinates $\textbf{r}_m = (x_m, y_m)^T$. We define $\mathcal{S}_m$ as the feasible set of all $\left(x, y\right)$ coordinates within its designated subarea. In other words, each $m$-th fluid element is capable of switching to one of $N_m=N_h^m\times N_v^m$ preset positions, where $N_h^m$ and $N_v^m$ are the numbers of preset positions per row and column within a 2D subarea $\mathcal{S}_m$. Due to its reconfigurable nature, each fluid element can dynamically adjust its position, phase shift, and power-splitting ratio to optimize both signal transmission and reflection. More precisely, to efficiently manipulate the incoming signals, each FIRES element controls two key properties: the power splitting and phase shifts. Following the energy splitting (ES) protocol, which offers significant flexibility for passive beamforming optimization, the power splitting ratio $\beta_{u,m}\in\left[0,1\right]$ of each $m$-th fluid element determines the proportion of the energy allocated to reflection ($\beta_{r,m}$) and transmission ($\beta_{t,m}$), ensuring that $\beta_{r,m} + \beta_{t,m} = 1$. To simplify the system design, we assume that all elements share identical amplitude coefficients \cite{ghadi2023analytical}, i.e., $\beta_{u,m}=\beta_u, \forall m$. 

Each $m$-th fluid element of the FIRES independently applies phase shifts to both the reflected and transmitted signals, denoted as $\phi_m^r$ and $\phi_m^t$, respectively, with $\phi_m^u\in\left(0,2\pi\right]$. As a result, the reflection and transmission coefficient matrices of the mteasurface can be expressed by $\mathbf{\Phi}_{u} = \mathrm{diag}\left(\sqrt{\beta_u} \mathrm{e}^{j \phi_1^u},\dots, \sqrt{\beta_u} \mathrm{e}^{j \phi_M^u}\right)$, where we use the notation $\Phi_{u}^{l}=\sqrt{\beta_u} \mathrm{e}^{j \phi_l^u}$ for each $l$-th entry, with $l=1,\ldots,M$, of the $\mathbf{\Phi}_{u}$ matrix. Therefore, for the ES protocol under multicast transmission \cite{mu2022sim}, the received signal at user $u$ is written as:
\begin{align}
y_u = \sqrt{P}\mathbf{h}_u^H \mathbf{\Phi}_u \mathbf{h}_f x + z_u,
\end{align}
where $P$ denotes the transmit power, $x$ is the common symbol transmitted to both users with unit power, and $z_u$ is the additive white Gaussian noise (AWGN) with zero mean and variance $\sigma^2$, i.e., $z_u\sim \mathcal{CN}\left(0,\sigma^2\right)$. The vector $\mathbf{h}_f \in \mathbb{C}^{M \times 1}$ represents the channel between the BS and the FIRES, whereas $\mathbf{h}_u \in \mathbb{C}^{M \times 1}$ is the channel between the FIRES and user $u$. As a result, the received signal-to-noise ratio (SNR) for user $u$ is given by the following expression:
\begin{align}
\gamma_u = \frac{P \left| \mathbf{h}_u^H \mathbf{\Phi}_u \mathbf{h}_f \right|^2}{\sigma^2}.\label{eq-snr}
\end{align}
%\subsection{Channel Model}
Due to the presence of a dominant line-of-sight (LoS) path and some scattered non-LoS (NLoS) components, the wireless channels in RIS-enabled systems are typically modeled as Rician fading channels~\cite{RIS_Ricean}. We follow this modeling in this paper, i.e., both the channels between the BS and FIRES, and between the FIRES and the users, exhibit Rician fading, accounting for the LoS component and the additional NLoS components resulting from reflections and diffraction. To this end, $\mathbf{h}_q$ for $q\in\{f,r,t\}$  is defined as follows:
\begin{align}
\mathbf{h}_q=\sqrt{l_q}\left(\sqrt{\frac{K_q}{K_q+1}}\mathbf{h}_{q,\mathrm{LoS}}+\sqrt{\frac{1}{K_q+1}}\mathbf{h}_{q,\mathrm{NLoS}}\right),
\end{align}
where $K_q$'s indicate the Rician factors for the BS-to-FIRES and FIRES-to-users links, and $\mathbf{h}_{q,\mathrm{LoS}}$'s are the LoS components, while $\mathbf{h}_{q,\mathrm{NLoS}}$'s denote the NLoS components, with $l_q$'s being the corresponding path-loss scaling factors. For the LoS components, we model the channels using the azimuth and elevation angle-of-departure (AoD) and angle-of-arrival (AoA). For the BS to FIRES link, the LoS component can be expressed as follows:
\begin{align}
\mathbf{h}_f = \mathbf{a}_f\left(\psi_{a_f}, \psi_{e_f}, \mathbf{r}\right) \mathbf{a}_b^H\left(\psi_b\right)
\end{align}
in which $\psi_b$ is the AoD from the BS and $a_b\left(\psi_b\right)$ denotes the transmit steering vector of the BS, which is defined as
$
\left[a_b\left(\psi_b\right)\right]_k = \mathrm{e}^{j\left(k-1\right)\pi \sin\left(\psi_b\right)}
$
with $[\cdot]_k$ denoting the $k$-th entry of the vector and $\psi_{a_f}$ and $ \psi_{e_f}$ being to the azimuth and elevation AoA at the FIRES, respectively. In addition, $\mathbf{a}_f\left(\psi_{a_f}, \psi_{e_f}, \mathbf{r}\right)$  is the receive steering vector at the FIRES which is defined as follows:
\begin{align}
\left[\mathbf{a}_f\left(\psi_{a_f}, \psi_{e_f}, \mathbf{r}\right)\right]_m=\mathrm{e}^{\frac{2\pi}{\lambda}\left(x_m\sin \psi_{a_f}\cos \psi_{e_f}+y_m\sin\psi_{e_f}\right)}.
\end{align}
Hence, the FIRES-to-user LoS channel is expressed as
\begin{align}
\mathbf{h}_u = \mathbf{a}_u\left(\psi_{a_u}, \psi_{e_u}, \mathbf{r}\right),
\end{align}
where $\psi_{a_u}$ and $\psi_{e_u}$ are the azimuth and elevation AoD from the FIRES to user $u$, respectively, and $\mathbf{a}_u\left(\psi_{a_u}, \psi_{e_u}, \mathbf{r}\right)$ is the steering vector for the FIRES to each user $u$, defined as:
\begin{align}
\left[\mathbf{a}_u\left(\psi_{a_u}, \psi_{e_u}, \mathbf{r}\right)\right]_m = \mathrm{e}^{\frac{2\pi}{\lambda}\left(x_m\sin \psi_{a_u}\cos \psi_{e_u}+y_m\sin\psi_{e_u}\right)}.
\end{align}

Given the possible small distance between adjacent preset positions, the spatial correlation between the FIRES elements needs to be considered. For the NLoS components, we model the spatial correlation using Jakes' model. Specifically, assuming any two arbitrary preset positions $\tilde{n}_q=\mathcal{F}^{-1}\left(\tilde{n}_h^q,\tilde{n}_v^q\right)$ and $\hat{n}_q=\mathcal{F}^{-1}\left(\hat{n}_h^q,\hat{n}_v^q\right)$, with $\mathcal{F}\left(\cdot\right)$ being a mapping function converting the 2D indices to the one-dimensional (1D) ones, the spatial correlation matrix takes the form \cite{ghadi2024on,Ricean_corr}
\begin{multline}\label{eq-rf}
\left[\mathbf{R}_q\right]_{\tilde{n}_q,\hat{n}_q}=\\\mathrm{sinc}\hspace{-1mm}\left(\frac{2}{\lambda}\sqrt{\left(\frac{\left|\tilde{n}_h^q-\hat{n}_h^q\right|}{L_h-1}A_h\right)^2+\left(\frac{\left|\tilde{n}_v^q-\hat{n}_v^q\right|}{L_v-1}A_v\right)^2}\right),
\end{multline}
where $\mathrm{sinc\left(t\right)=\frac{\sin\left(\pi t\right)}{\pi t}}$. The terms $L_h$ and $L_v$ indicate the number of preset positions per row and column, respectively, of the entire meatsurface, such that $L=L_h\times L_v = \sum_{m=1}^M N_m$ is the total number of preset positions. The spatial correlation matrix based on \eqref{eq-rf} can be further written as $\mathbf{R}_q = \mathbf{U}_q \Lambda_q \mathbf{U}_q^H$, 
where $\mathbf{U}_q$ denotes the unitary matrix containing $\mathbf{R}_q$'s eigenvectors and $\mathbf{\Lambda}_q$ is the diagonal matrix containing its eigenvalues. Consequently, the NLoS component is expressed as $\mathbf{h}_{q,\mathrm{NLoS}} = \mathbf{\Lambda}_q^{1/2}\mathbf{U}_q^{1/2}\overline{\mathbf{h}}_q$, where $\overline{\mathbf{h}}_q$ represents the uncorrelated small-scale fading component, modeled as independent and identically distributed (i.i.d.)~complex Gaussian variables with zero mean and unit variance.

\vspace{-2mm}
\section{Problem Formulation}\label{sec-main}
We focus on the maximization of the effective rate of the proposed FIRES-enabled communication system. The achievable rate for user $u$ is computed as %\begin{align}
$R_u=\log_2\left(1+\gamma_u\right)$.
%\end{align}
Consequently, the effective rate under the ES protocol is defined as $R_\mathrm{eff} \triangleq \min \{R_r, R_t\}$ \cite{mu2022sim}. To reach an optimal effective rate, each fluid FIRES element must be assigned a phase shift that precisely counterbalances the total phase shift accumulated over both the first and second hops, i.e.: 
\begin{align}
\Phi_u^{m*}=\sqrt{\beta_u}\mathrm{e}^{-j\left(\angle\left[\mathbf{h}_f\right]_m+\angle\left[\mathbf{h}_u\right]_m\right)}, 
\end{align}
and thus, the achievable rate for user $u$ is rewritten as follows:
\begin{align}
R_u = \log_2\left(1+\frac{P\left|\sum_{m=1}^M\left|\left[\mathbf{h}_{f,m}\right]\right|\left|\left[\mathbf{h}_{u,m}\right]\right|\right|^2}{\sigma^2}\right).
\end{align}
Hence, our design objective is the maximization of the minimum of the achievable rates for both the transmitting and reflecting users under the system constraints; in particular: 
\begin{subequations}
	\begin{align}
		\left(\mathrm{P1}\right):&\,\, \underset{\mathbf{r}}{\max}{\,\,R_\mathrm{eff}}\\ \label{eq-rm}
		&\,\,\mathrm{s.t.}\quad \mathbf{r}_m\in \mathcal{S}_m, \forall m,\\ \label{eq-sm}
		&\,\,\quad\quad  \|\mathbf{r}_m-\mathbf{r}_{m'}\|_2\ge D, \forall m\neq m',\\
		&\,\,\quad\quad \, \,\beta_r+\beta_t=1,\\
		&\,\,\quad \quad \, \, P_r+P_t\leq P,
	\end{align}
\end{subequations}
where $\mathbf{r}=\left[\mathbf{r}_1,\dots,\mathbf{r}_M\right]$ and $D$  indicates the minimum spacing set between two fluid elements. Note that mutual coupling between fluid elements in FIRES can be neglected, as the minimum spacing $D$ can be carefully determined by the chosen FAS technology to significantly reduce such effects. In reconfigurable or radio-frequency pixel-based metasurfaces, dynamic control over element positions, phase, and power splitting minimizes coupling, rendering its impact negligible for the scope of this work~\cite{chiu2007reduction}.

Due to the non-convex nature of the optimization problem $(\mathrm{P1})$, conventional optimization techniques may struggle to find the global optimal solution. To overcome this challenge, we employ a PSO approach, which is well-suited for solving high-dimensional, non-convex problems like $(\mathrm{P1})$. In the considered FIRES system, each particle in the swarm represents a candidate solution, where the positions of the fluid elements is encoded in the particle's position vector. The velocity of each particle represents the direction and magnitude of movement within the solution space. PSO iteratively updates the positions and velocities of particles based on two factors, namely, the particle's own best position and the best position found by any particle in the swarm. In particular, the velocity is updated as:
\begin{align}
\mathbf{v}_i^{\left(t+1\right)} = w\mathbf{v}_i^{(t)}+c_1r_1\left(\mathbf{r}_{i,\mathrm{best}}-\mathbf{r}_i^{(t)}\right)+c_2r_2\left(\mathbf{r}_\mathrm{best}-\mathbf{r}_i^{\left(t\right)}\right),
\end{align}
in which $w$ is the inertia weight that controls the balance between exploration and exploitation, $c_1$ and $c_2$ are the cognitive and social learning factors, $r_1$ and $r_2$ are random numbers between $0$ and $1$, $\mathbf{r}_{i,\mathrm{best}}$ is the best position found by the individual particle, and $\mathbf{r}_{\mathrm{best}}$ is the global best position found by the swarm. The position of each particle is updated as:
\begin{align}
\mathbf{r}_i^{(t+1)} = \mathcal{P}\left(\mathbf{r}^{(t)}_i+\mathbf{v}_i^{\left(t+1\right)}\right),
\end{align}
where $\mathcal{P}\left(\cdot\right)$ is the function enforcing the constraint \eqref{eq-sm}.

Incorporating the constraints into the optimization problem requires modifying the objective function by introducing penalty terms. These penalties ensure that the positions of the FIRES fluid elements satisfy the minimum spacing and power budget constraints. For the power budget constraint, we define a penalty term that is added to the fitness function whenever the sum of $P_t + P_r$ exceeds the power budget $P$, i.e.:
\begin{align}
\mathfrak{B}_\mathrm{power}=\max\left\{0,P_t+P_r-P\right\}.
\end{align}
Moreover, for the minimum spacing constraint, a penalty term is added when the distance between any two fluid elements falls below the minimum spacing $D$ so that:
\begin{align}
\mathfrak{B}_\mathrm{spacing} = \sum_{1 \leq m < m' \leq M}{\bf 1}\left[\|\mathbf{r}_m-\mathbf{r}_{m'}\|<D\right].
\end{align}
Thus, the total penalty function is the sum of these individual penalties, and the objective function is modified as follows:
\begin{multline}
O\left(\mathbf{r}_i^{(t)}\right) =\sum_{u\in\left\{r,t\right\}}R_u\left(\mathbf{r}_i^{(t)}\right)\\
-\tau\left(\mathfrak{B}_\mathrm{power}\left(\mathbf{r}_i^{(t)}\right)+\mathfrak{B}_\mathrm{spacing}\left(\mathbf{r}_i^{(t)}\right)\right),
\end{multline}
where $\tau$ is a large positive penalty coefficient that ensures the constraints are strictly enforced.

The PSO algorithm proceeds iteratively, with particles exploring the solution space and updating their positions and velocities based on the fitness evaluations and the penalty function. The optimization process continues until the algorithm converges, at a point where the positions of the FIRES fluid elements maximize the effective rate, while the system constraints (i.e., power budget, power splitting, and inter-element spacing) are properly met. The proposed PSO-based algorithm solving $(\mathrm{P1})$ is summarized in Algorithm~\ref{algo1}. 
%Algorithm~\ref{algo1} summarizes the proposed PSO-based optimization framework, which iteratively adjusts the positions of the fluid elements to maximize the achievable effective rate under spatial spacing constraints.
\begin{algorithm}[t]
\caption{PSO-Based Optimization for FIRES}
\label{algo1}
\begin{algorithmic}[1]
\STATE \textbf{Input:} \(A = [X_{\min}, X_{\max}] \times [Y_{\min}, Y_{\max}]\), \(M\), \(D\), \(P\), \(T\), \(w\), and \(c_1, c_2\).
\STATE \textbf{Initialization:} Partition \(A\) into valid $M$ subareas with spacing \(\geq D\); precompute \(\mathbf{h}, \mathbf{h}_r\), and \(\mathbf{h}_t\); initialize correlation matrices \(\mathbf{J}\).
\FOR{each FIRES fluid element \(m = 1 \text{ to } M\)}
    \STATE Randomly initialize \(\mathbf{r}_m^p\) and \(\mathbf{v}_m^p\).
    \FOR{each iteration \(t = 1 \text{ to } T\)}
        \STATE Update \(\mathbf{J}_m^p\) for each particle based on \(\mathbf{r}_m^p\).
        \STATE Compute SNRs \(\gamma_t\) and \(\gamma_r\) using \eqref{eq-snr}.
        \STATE Evaluate fitness \(R_m^p = \min \{\log_2(1 + \gamma_t), \log_2(1 + \gamma_r)\)\}.
        \STATE Update personal best \(\mathbf{p}_m^p\) and global best \(\mathbf{h}_m\).
        \STATE Update velocities as \(\mathbf{v}_m^p \leftarrow w\mathbf{v}_m^p + c_1 r_1 (\mathbf{p}_m^p - \mathbf{r}_m^p) + c_2 r_2 (\mathbf{h}_m - \mathbf{r}_m^p)\).
        \STATE Update positions as \(\mathbf{r}_m^p \leftarrow \mathbf{r}_m^p + \mathbf{v}_m^p\), applying boundary constraints.
    \ENDFOR
    \STATE Select \(\mathbf{r}_m^*\) as the best position of element \(m\).
    \IF{any \(\|\mathbf{r}_m^* - \mathbf{r}_{m'}^*\| < D\) for \(m' < m\)}
        \STATE Re-optimize \(\mathbf{r}_m^*\) within subarea to satisfy spacing.
    \ENDIF
    \STATE Append \(\mathbf{r}_m^*\) to output set,
\ENDFOR
\STATE \textbf{Output:} Final \(\{\mathbf{r}_m^*\}_{m=1}^M\) and corresponding \(R_u\)'s.
\end{algorithmic}
\end{algorithm}

\vspace{-2mm}
\section{Simulation Results and Discussion}\label{sec-sim}
In this section, we present simulation results to evaluate the performance of the proposed FIRES-aided system in terms of the effective rate performance. Unless otherwise specified, the simulation parameters are configured as follows: $M = \{4,9,16\}$, $ N = 100$, $A = 4$ m$^2$, $\sigma^2 = -90$ dBm, $K_f = 5$, $K_u = 5$, $t = 100$, $i = 50$, $w = 0.4$ $c_1 = 0.5$, and $c_2 = 0.5$. We denote the distance between the BS and FIRES as $d_f = 100$ m, and the distance between FIRES and the users as $d_u = 200$ m. The path-loss exponent has been set as $\alpha=2.5$. Additionally, we assume a minimum element spacing of $D=\lambda/2$, with the carrier frequency $f_c$ set to $3.5$ GHz. The AoA and AoD at both the BS and FIRES are independently and uniformly distributed over the interval $\left(0,\pi\right).$ Moreover, we benchmark the performance against a conventional STAR-RIS-enabled communication system, where the metasurface has the same size as the FIRES with $\widehat{M}$ denoting the total number of its tunable elements. 

\begin{figure*}[htbp]
	\centering
	\subfigure[]{
		\includegraphics[width=0.33\linewidth]{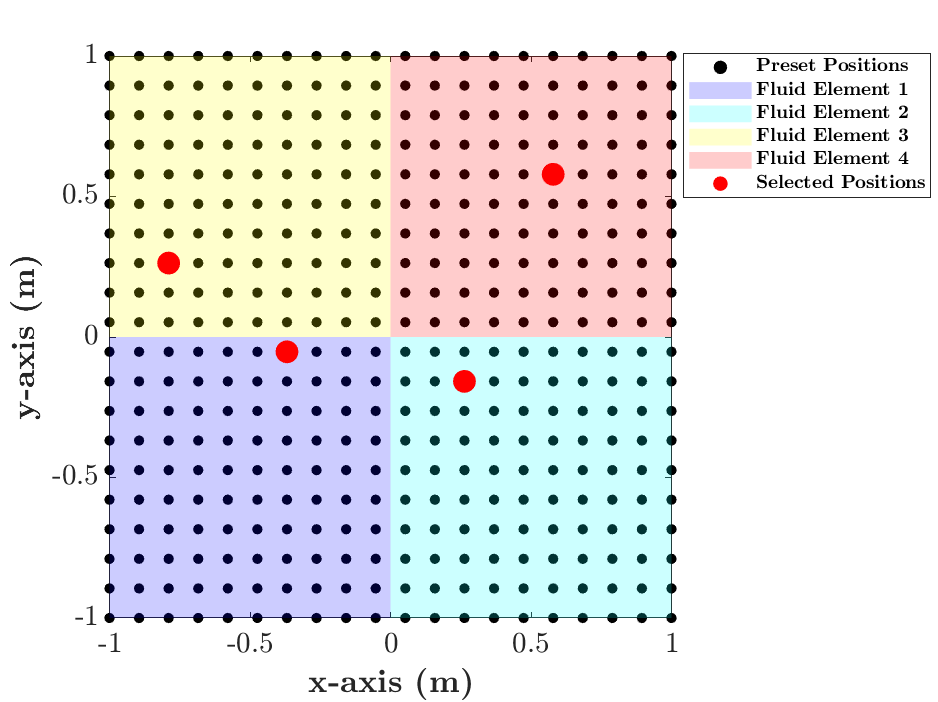}
		\label{fig_4}
	}\hspace{-0.9cm}
	\hfill
	\subfigure[]{
		\includegraphics[width=0.33\linewidth]{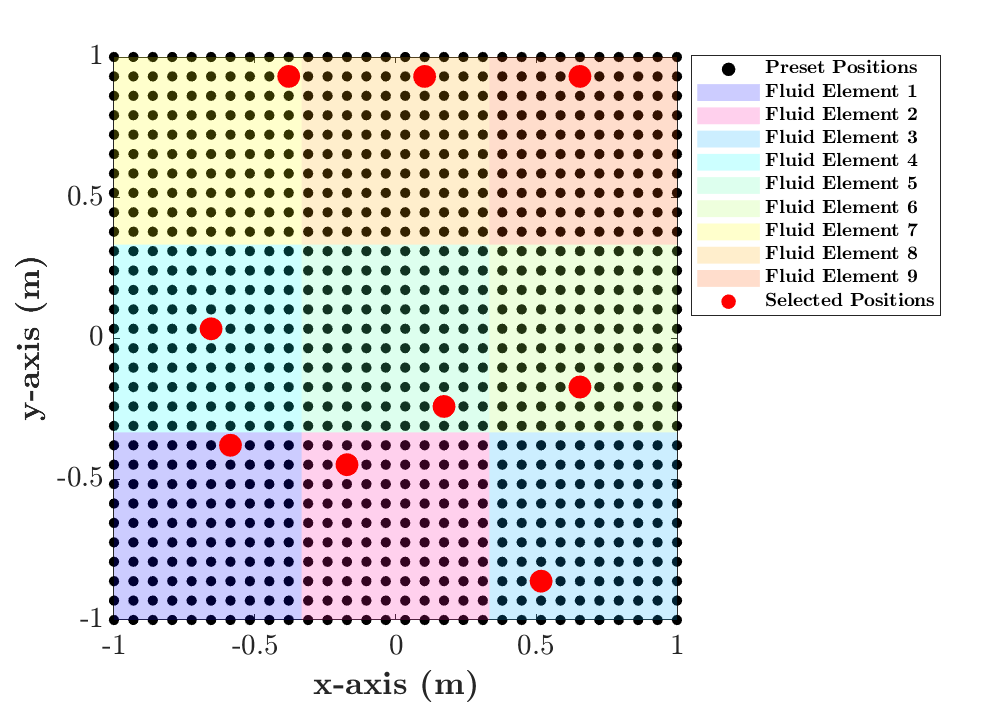}
		\label{fig_9}
	}\hspace{-0.5cm}
    	\subfigure[]{
		\includegraphics[width=0.33\linewidth]{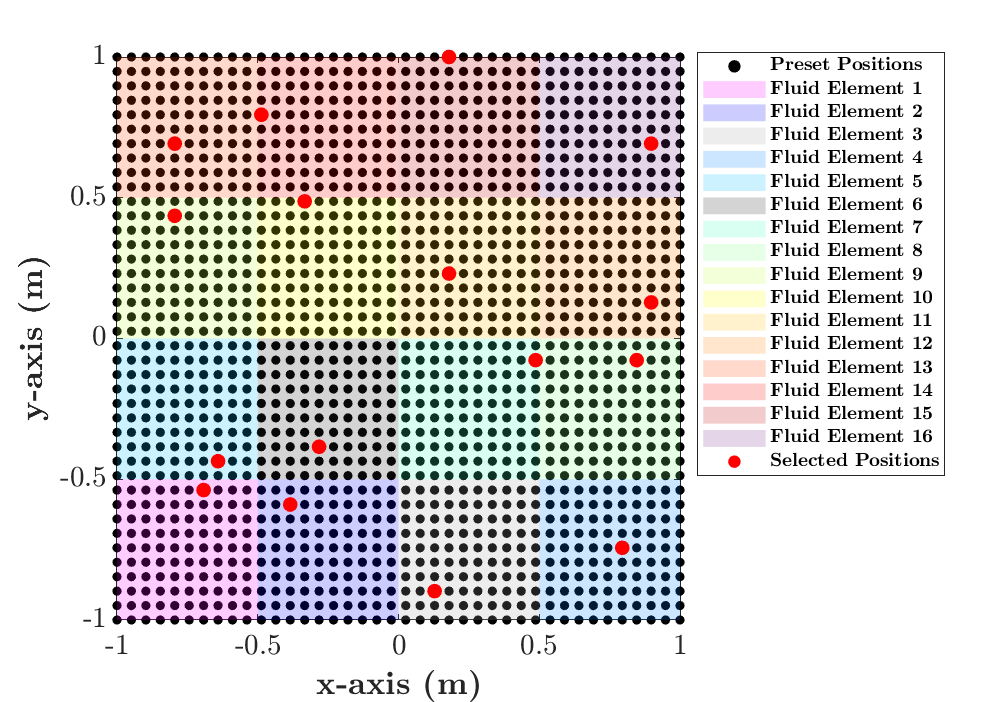}
		\label{fig_16}
	}\vspace{-0.3cm}
	\caption{Illustration of examples of the optimized positions when the FIRES has (a) $M=4$; (b) $M=9$; and (c) $M=16$ fluid elements.}\vspace{-0.5cm}
	\label{fig_loc}
\end{figure*}

\begin{figure*}[htbp]
	\centering
	\subfigure[]{
		\includegraphics[width=0.33\linewidth]{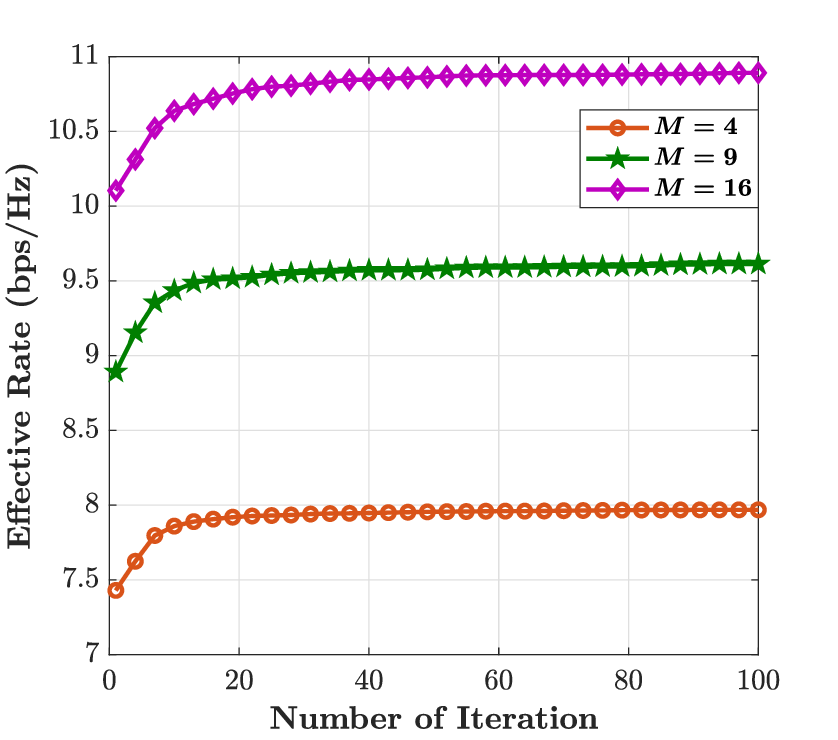}
		\label{fig_it}
	}\hspace{-0.9cm}
	\hfill
	\subfigure[]{
		\includegraphics[width=0.33\linewidth]{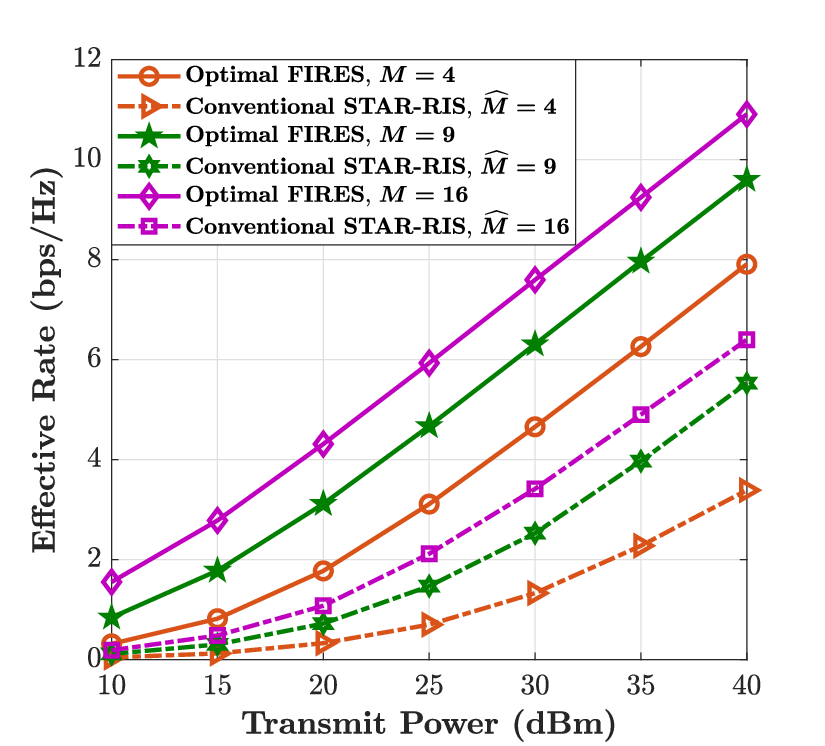}
		\label{fig_n}
	}\hspace{-0.5cm}
    	\subfigure[]{
		\includegraphics[width=0.33\linewidth]{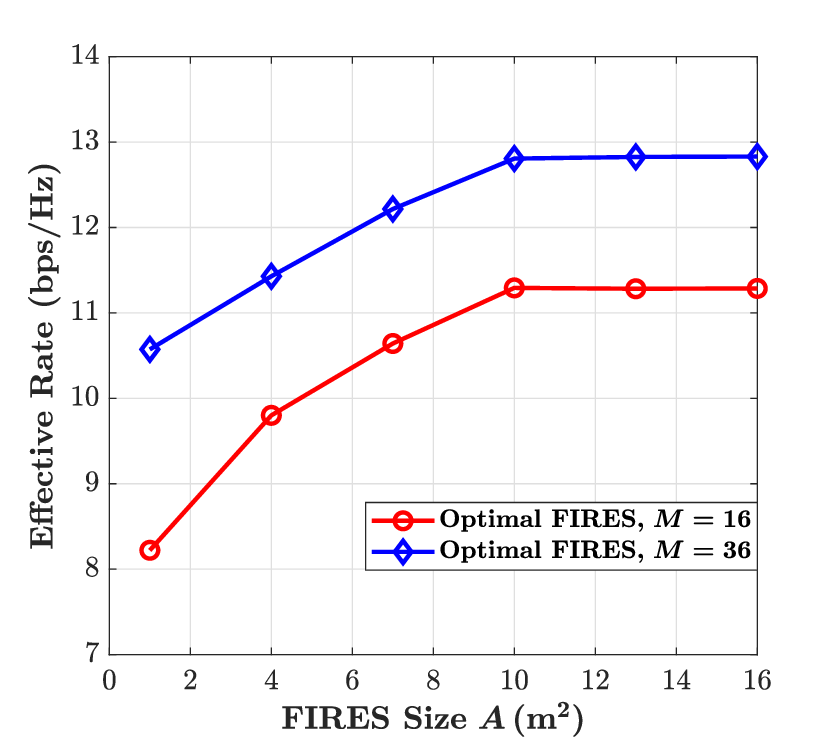}
		\label{fig_a}
	}\vspace{-0.3cm}
\caption{Effective rate versus: (a) number of iterations of Algorithm~\ref{algo1} for different numbers $M$ of FIRES fluid elements when $P=40$ dBm; (b) transmit power $P$ for different $M$; and (c) size of FIRES $A$ for different $M$ when $P=40$ dBm and $N_h^m=N_v^m=100$.}\label{fig-rate}
\end{figure*}

Figures~\ref{fig_4}--\ref{fig_16} illustrates the preset location options for the FIRES with $M=4$, $9$, and $16$ fluid elements, respectively. For each fluid element (subarea), the most favorable preset is chosen based on the considered optimization framework. The selected positions are spread across the subareas in a way that supports performance improvement. Nevertheless, the connection between the preset configuration and the achievable rate is intricate and varies with the instantaneous channel conditions. Figure~\ref{fig-rate} assesses the effective rate performance for different scenarios. The convergence behavior of the proposed PSO-based optimization algorithm for different values of $M$is demonstrated in Fig.~\ref{fig_it}. Efficient convergence is showcased in all cases, with most rate improvements occurring during the initial iterations. This highlights the algorithm's strong ability to exploit favorable regions of the search space. Nevertheless, as $M$ increases, convergence becomes more gradual due to the increased dimensionality of the optimization problem. As expected, the higher number of fluid elements introduces additional DoF, requiring more iterations to reach optimal configurations. These results confirm that, while the algorithm remains effective across all configurations, larger FIRES setups necessitate more computational effort to achieve full optimization.

Figure.~\ref{fig_n} compares the effective rate performance of the proposed FIRES with that of the conventional STAR-RIS for different $M=\widehat{M}\in\{4,9,16\}$ values. The results indicate that, increasing the number of elements, leads to significant rate improvements for both schemes. More perceptibly, we can observe that FIRES consistently outperforms its conventional counterpart in both modes, due to the fluid elements' ability to dynamically adapt their positions, enhancing signal paths and optimizing propagation conditions. For instance, at $35$ dBm transmit power, FIRES achieves approximately $8$ bps/Hz, while conventional STAR-RIS attains about $4$ bps/Hz, resulting in a performance improvement of nearly twofold. These improvements stem from the superior spatial flexibility of FIRES, which enhances multi-user communications by providing more favorable signal paths and increased constructive interference.

The impact of the FIRES size $A$ on the effective rate performance is illustrated in Fig.~\ref{fig_a}. By assuming a fixed transmit power of $P=40$ dBm, we see that increasing $A$ leads to higher effective rate values, especially when the grid resolution is high. This improvement arises because, enlarging the overall radiating surface reduces the spatial correlation among fluid elements, thereby enhancing spatial diversity. As a result, the channel becomes more decorrelated and distinct, allowing the system to better exploit the available propagation paths and improve overall performance. However, it is important to note that this gain becomes more pronounced when the grid resolution in the FIRES is sufficiently high, as a finer grid allows the system to better sample the spatial variations across the FIRES. For instance, at $M=16$, increasing the FIRES size from $1\,\mathrm{m}^2$ to $16\,\mathrm{m}^2$ leads to improvements of $37\%$ in the effective rate. Furthermore, a saturation effect is observed when increasing the FIRES size while keeping the number of fluid elements $M$ fixed. This is because the spatial density of fluid elements decreases, limiting the system's ability to fully exploit the enlarged aperture. Also, considering a larger $M$ while increasing the size $A$ leads to further improvements in the effective rate, as the system benefits from both a larger radiating surface and higher spatial resolution. Nevertheless, even under this joint scaling, the rate enhancement eventually saturates due to fundamental limitations in channel richness and the finite number of spatial DoF available. 

\section{Conclusion}\label{sec-con}
This letter proposed the FIRES concept, a novel fluid-based STAR-RIS architecture that integrates both electromagnetic and geometric reconfigurability to faciliate highly adaptive smart wireless environments. Given each `fluid' element's capability of simultaneously reflecting and transmiting signals while also dynamically repositioning in space, FIRES introduces a new level of physical agility and signal control. To fully exploit its potential under the ES protocol, we formulated a non-convex effective rate maximization problem, jointly optimizing spatial positions, transmission-reflection coefficients, and operating modes. This was effectively addressed using a PSO algorithm tailored to the problem's high-dimensional and non-linear nature. Our simulation results demonstrated that FIRES consistently outperforms conventional STAR-RISs, establishing its promise for next-generation wireless networks.

\end{document}